\documentclass[12pt]{article}
 \immediate\openin0=hyperref.sty
 \ifeof0\immediate\closein0\else\immediate\closein0\usepackage[hyperindex]{hyperref}\fi
 \let\oldensuremath=\ensuremath\renewcommand\ensuremath{\relax\oldensuremath}

\newcommand\units[1]{\ensuremath{\mathbin{}{\rm#1}}}
\newcommand\MeV{\units{MeV}}
\newcommand\GeV{\units{GeV}}
\def\etal{{\it et al.}}

\newcommand{\be}{\begin{equation}}
\newcommand{\ee}{\end{equation}}
\newcommand{\ba}{\begin{eqnarray}}
\newcommand{\ea}{\end{eqnarray}}

\newcommand{\MSbar}{{\overline{MS}}}
\newcommand{\gsim}{\ensuremath{\mathrel{\raise2pt\hbox to 8pt{%
			\raise -5pt\hbox{$\sim$}\hss{$>$}}}}}

 \newif\ifchi\newif\iflat\newif\ifMeV
 \newif\iflight\newif\ifinter\newif\ifstrange
 \newif\ifK\newif\ifKstar\newif\ifphi
 \chifalse\lattrue\MeVtrue \lighttrue\intertrue\strangetrue
 \Ktrue\Kstartrue\phifalse

 \newcount\colcount\newcount\allcolcount
 \newcommand\setupcounts{%
    \colcount=0
    \iflight  \advance\colcount by 1\fi
    \ifinter  \ifK    \advance\colcount by 1\fi
              \ifKstar\advance\colcount by 1\fi
              \ifphi  \advance\colcount by 1\fi\fi
    \ifstrange\ifK    \advance\colcount by 1\fi
              \ifKstar\advance\colcount by 1\fi
              \ifphi  \advance\colcount by 1\fi\fi
    \allcolcount=1
    \ifchi\advance\allcolcount by 1\fi
    \iflat\advance\allcolcount by \colcount\fi
    \ifMeV\advance\allcolcount by \colcount\fi
    \ifcase\colcount\begingroup\errhelp{can't switch off all columns}%
                               \errmessage{colcount is 0}\endgroup\fi
 }

\input{epsf}

\begin{document}

\title{ \vspace{-1cm}\begin{flushright}
{\small LAUR-99-4581 \\}
\end{flushright}
{
Prospects of calculating $\epsilon_K$ and $\epsilon'$ from lattice QCD
}}

\author{
{\bf Rajan Gupta}\\[.6cm]
\small Theoretical Division, MS-B285, Los Alamos National Laboratory, \\
\small Los Alamos, NM 87545, USA.\\[.2cm]
}

\maketitle

\begin{abstract}
\noindent
An overview of lattice results for the strange quark mass, $B_K$, 
$B_6$, $B_7$, and $B_8$ is presented. I give my assessment of the 
reliability of various estimates and prospects for future 
improvements. 

\vspace{.1in}

\end{abstract}

\section{Introduction}
\label{sec:intro}

The status of CP violation in Kaon decays is 
\begin{eqnarray}
\epsilon                     &=& (2.280 \pm 0.013) 10^{-3} e^{i \pi/4}   \nonumber \\
{\rm Re}(\epsilon'/\epsilon) &=& (21.2 \pm 4.6) 10^{-4}
\end{eqnarray}
where I have taken the most recent world average of
$\epsilon'/\epsilon$ from M. Sozzi's talk. The experimental errors on
$\epsilon'/\epsilon$ will decrease once KTeV and NA48 collaborations
analyze their full data set, providing us with a unique opportunity to
test the standard model.  The spotlight is, therefore, now on theory:
can one reconcile the two measurements with the predictions of the
standard model in which both parameters are governed by the single
phase in the CKM matrix, or are these results a window to new physics?

The standard model predictions, evaluated using the effective weak
Hamiltonian defined at scale $\mu$, are summarized by Buras in his talk:
\begin{eqnarray}
\epsilon &=&    
     {\rm Im}\lambda_t C_\epsilon {\hat B_K}  e^{i\pi/4}
    \big\{\, {\rm Re} \lambda_c \left[ \eta_1 S_0(x_c) - \eta_3 S_0(x_c,x_t) \right]  + 
    \ {\rm Re}  \lambda_t \eta_2 S_0(x_t) 
     \ \big\} \nonumber \\
\epsilon' &=& \frac{ie^{i(\pi/4 + \delta_2 - \delta_0)}}{\sqrt2}\ \frac{{\rm Im} A_2}{{\rm Re} A_0} 
            \left[1 - \epsilon^2 \right] \nonumber \\
          &=& {\rm Im}\lambda_t \frac{G_F e^{i(\pi/4+\delta_2 - \delta_0)}}{2 {\rm Re} A_0}\
            \left[ \omega \sum_{i} y_i \langle Q_i \rangle_0 ( 1 - \Omega_{\eta + \eta'}) - 
                          \sum_{i} y_i \langle Q_i \rangle_2 \right]
\end{eqnarray}
where $C_\epsilon = 3.78 \times 10^4$, $\omega = {\rm Re} A_2 / {\rm
Re} A_0 \approx 1/22$, $\Omega_{\eta + \eta'}$ is the isospin breaking contribution, 
$y_i$ are the Wilson coefficients, $\langle Q_i
\rangle_I = \langle (\pi\pi)_I | Q_i | K \rangle$, and the sum is over
all the 4-fermion operators that contribute. Also, I use the
convention ${\rm Im} A_0 = 0$ and in the last expression neglect the term proportional to 
$\epsilon^2$. As expected, both quantities are
proportional to ${\rm Im} \lambda_t \equiv {\rm Im} V_{td}V_{ts}^* =
A^2 \lambda^5 \eta$ in the Wolfenstein parameterization.

The equation for $\epsilon$ provides a parabolic constraint in the
$\rho - \eta$ plane provided $\hat B_K$ and $|V_{cb}|$ are
known. Alternately, a precise determination of $\hat B_K$ and
$|V_{cb}|$ would fix ${\rm Im} \lambda_t$, and the measurements of
$\epsilon'/\epsilon$ could be used to look for new physics. Note that
a larger value of $\hat B_K$ implies smaller ${\rm Im} \lambda_t$, and
consequently smaller $\epsilon'$.

For $\epsilon'$ we follow the work of Buras
\etal\ \cite{CP96Buras} who use the relations between the various $\langle Q_i \rangle$
and make maximal use of experimental input. Then
\begin{equation}
\left| \frac{\epsilon'}{\epsilon} \right| = {\rm Im}\lambda_t \bigg\{ 
	c_0 + \big( c_6 B_6^{1/2} + c_8 B_8^{3/2} \big) 
              \left( \frac{158 \MeV}{m_d+m_s} \right)^2 \bigg\}
\label{eq:epmaster}
\end{equation}
For $\mu = m_c$, $\Lambda_{QCD}^{(4)} = 325 MeV$, and central values
of the other parameters, Buras \etal\ \cite{CP96Buras} get in the NDR
scheme
\begin{eqnarray}
{\rm Im}\lambda_t  &\approx& 1.29 \times 10^{-4} \,, \nonumber\\
c_0  &\approx& -1.4 \,, \nonumber\\
c_6  &\approx& +7.9 \,, \nonumber\\
c_8  &\approx& -4.1 \,.
\label{eq:epmasternum}
\end{eqnarray}
From Eqs.~\ref{eq:epmaster},\ref{eq:epmasternum} it is clear that
there is a strong cancellation between the $\Delta I = 1/2$ (dominated
by QCD penguin $Q_6$) and $\Delta I = 3/2$ (dominated by electro-weak
penguin $Q_8$) operators, and the value of the strange quark mass
plays a crucial role.  For $B_6 = B_8 = 1$ (vacuum saturation
approximation (VSA) values currently used as inputs), one needs $m_s +
m_d = 70$ MeV at $\mu = m_c$ instead of 158 MeV to get $\epsilon' /
\epsilon $ to $\approx 23 \times 10^{-4}$.  A more likely scenerio is
an enhancement of $B_6$, a suppression of $B_8$ and the quark masses,
and/or a conspiracy of all other input parameters. It is therefore
important to determine all three quantities, $m_s$, $B_6$, and $B_8$,
accurately in order to resolve whether or not the SM predicts the
observed value of $\epsilon'$.

There are two omissions from my subsequent discussion of lattice
calculations. First, I defer to T. Blum's talk for lattice results
using domain wall fermions (DWF). Second, G. Martinelli has proposed
analyzing $B$-parameters without introducing a dependence on
$m_s+m_d$. Recall that the dependence on $m_s+m_d$ is introduced
because in VSA $\langle O_6 \rangle, \langle O_8 \rangle \propto
|\langle 0 | {\cal P} | K \rangle|^2 = 4 M_K^4 f_K^2 / (m_s +
m_d)^2$. Using $B$ parameters as commonly defined has certain
numerical advantages, but it does shift the scale dependence of the
operator into a new quantity $(m_s + m_d)$. Which approach is better is a
matter of numerical detail, and since at this point I do not have data
to make comparative statements, I direct the reader to Martinelli's 
writeup.

\section{$B_K$}

Even though most calculations of $B_K$ have been done in quenched QCD,
there are good reasons to believe, as discussed below, that we have a
reasonable estimate for the full theory.  A summary of results is
presented in Table \ref{t:BK}.  The most precise calculation in terms
of both statistical and systematic errors, is by the JLQCD
collaboration. Their result $B_K = 0.628(42) $ when converted to the
renormalization group invariant ${\hat B}_K$ is~\cite{CPMASS97}
\begin{equation}
{\hat B}_K = 0.86(6) \,.
\label{eq:BK}
\end{equation}
Since this result is for the quenched theory we have to address two
associated issues. (i) Quenched chiral logs (QCL), and (ii) other
effects of quenching. The other remaining systematic error is the use
of degenerate quarks for the kaon. The quark mass is typically varied
in the range $3 m_s - m_s/3$, and the physical kaon is defined in the
following two ways. (a) It is assumed to be made up of
two degenerate quarks of mass $m_s/2$, or (b) the ``light'' quark is
extrapolated to $m_d$ and the other interpolated to $m_s$. The issue
of quenched chiral logs is therefore relevant to (b).  (The reason is
that for degenerate quarks, quenched QCD and QCD have the same chiral
expansion and enhanced logs due to $\eta'$, an additional Goldstone
boson in the quenched theory, vanish~\cite{sharpeTASI94}.) The status
on each of these issues and some clarifications on the different
published numbers is as follows.

\begin{table}
\begin{center}
\medskip\noindent
\begin{tabular}{|l|c|c|c|}
\hline
    Fermion Action                     &  $Z$       &  $\beta$     &  $B_K(\overline{MS},2 \GeV)$ \\
\hline							              
Staggered \cite{B97stag}           & 1-loop         &  $6.0-6.4^*$   &  0.62(2)(2) \\
Staggered \cite{BK98stagJLQCD}     & 1-loop         &  $5.7-6.65^*$  &  0.628(42) \\
Staggered \cite{BK97stagGREG}      & 1-loop         &  $5.7-6.2^*$   &  0.573(15) \\
								     
Wilson    \cite{B97wilnLANL}       & 1-loop TI      &  $6.0$       &   0.74(4)(5) \\
Wilson    \cite{BK99wilnJLQCD}     & 1-loop \& $\chi WI$ &  $5.9-6.5^*$   &   0.69(7) \\
Clover \cite{BKB7B897APE}          & 1-loop BPT     &  $6.0$       &   0.65(11) \\
Clover \cite{BKB7B897APE}          & Non-pert.      &  $6.0$       &   0.66(11) \\
TI Clover \cite{B98lellouch}       & 1-loop TI      &  $6.0,6.2$   &   0.72(8) \\
\hline
\end{tabular}
\caption{Lattice estimates of $B_K(NDR,\mu=2\GeV)$ for different
lattice actions. An asterisk imples that the data were extrapolated to
$a=0$.}
\label{t:BK}
\smallskip
\end{center}
\end{table}

\begin{itemize}

\item
The parity even part of the 4-fermion operator has two terms, $VV$ and
$AA$.  Sharpe~\cite{sharpePRD92,sharpeTASI94} has calculated the QCL in these and
their lattice volume dependence.  The lattice data show the expected
behavior, providing a basis for confidence in the CPT analyses.
The leading chiral log cancels in the sum, $VV + AA$, thus alleviating
the uncertainties associated with chiral extrapolations in the
quenched theory.
\item
Estimates of quenching uncertainties provided by CPT are 
strengthened by the preliminary unquenched calculations~\cite{kilcupBKfull}, 
and suggest that $B_K({\rm full\ QCD}) \approx 1.05 B_K({\rm quenched})$.
\item
CPT has also been used by Sharpe~\cite{sharpeTASI94} to estimate the
uncertainty associated with using ($m_d \approx m_s$) rather than the
physical ratio ($m_d \approx 0.055 m_s$). He estimates $B_K({\rm
QCD}) \approx 1.05 \pm 0.05 B_K({\rm degenerate})$.
\item
Lastly, it is a fortunate coincidence that the conversion of quenched
$B_K(\overline{MS}, \mu =2 \GeV)$ to ${\hat B}_K$ is very insensitive
to whether one uses quenched $\alpha_s$ and anomalous dimensions or
those for the full theory.
\end{itemize}

The success of CPT in estimating errors raises the question whether
the systematic shifts due to quenching and degenerate versus physical
mass quarks discussed above should be incorporated in the final value
of ${\hat B}_K$ or stated as a separate error? Sharpe in
\cite{BK98vanc} includes them and quotes ${\hat B}_K = 0.94 \pm 0.06
\pm 0.14$, where the second error is a very conservative estimate of
the systematic uncertainties. In \cite{CPMASS97}, I chose not to
include them in the central value and had used a more aggressive
estimate of systematic errors in quoting ${\hat B}_K = 0.86 \pm 0.06
\pm 0.06$. Both estimates are based on exactly the same data (JLQCD),
and until unquenched data of comparable quality becomes available the
choice between them reflects one's taste in the handling of systematic
errors.

Finally, in my view, one should not average the numbers given in
Table~\ref{t:BK} to get a ``best'' lattice result. At present, JLQCD's
is, by far, the best calculation with respect to lattice size,
statistics, and systematics. (The quoted errors in Table~\ref{t:BK} do
not always include/address all the systematics uncertainties equally
well).  What the table does highlight is that all the results agree: a
confirmation of the stability of lattice calculations of $B_K$.

\section{Light quark masses}  

Since mid-1996 there has been a spurt of activity in the calculation
of light quark masses from both lattice QCD and QCD sum-rules.  The
intriguing possibility first suggested by lattice calculations is that
$m_u$, $m_d$, and $m_s$ are much lighter than previous estimates based
on QCD sum-rules. For a summary of the lattice methodology and results 
until Oct. 1997 see \cite{Mqrev97} and also the talk by S. Ryan \cite{Mq99ryan}. 

Recent quenched lattice results are summarized in Table~\ref{t:mq} and
unlike $B_K$ there is no single calculation that is ``superior'' to
the rest. (Unfortunately, once again this is not obvious from quoted
errors.) At first glance one sees a significant spread. Focusing
attention on $m_s$ extracted by fixing $M_K$ to its physical value,
$m_s(M_K)$, the estimates lie between $90-115$ MeV. A large part of
this variation is due to the quantity used to set the lattice scale
$1/a$. There is also a large difference between $m_s(M_K)$ and
$m_s(M_\phi)$, $i.e.$ different strange mesons give different
estimates; and even though neither one reproduces the octet and
decuplet baryon mass splittings, $m_s(M_\phi)$, comes much closer
\cite{HM96LANL,MqCPPACS99}.  A short explanation of the results is in
order.

\begin{table}
\begin{center}
\medskip\noindent
\begin{tabular}{|l|l|c|c|c|c|}
\hline
                                  &  Action    & $\bar m$        & $m_s(M_K)$  & $m_s(M_\phi)$& scale $1/a$    \\
\hline
Summary 1997 \cite{Mqrev97}       &            & $3.8(1)(3)$     & $99(3)(8)$  & $111(7)(20)$ & $M_\rho$       \\
APE     1998 \cite{MqAPE98}       & O(a) SW    & $4.5(4)$        & $111(12)$   &              & $M_{K^*}$    \\
APE     1999 \cite{MqAPE99}       & O(a) SW    & $4.1(6)$        & $98(12) $   &              & $M_{K^*}$    \\
CPPACS  1999 \cite{MqCPPACS99}    &  Wilson    & $4.55(18)$      & $115(2)$    & $143(6)$     & $M_\rho$       \\
JLQCD   1999 \cite{MqJLQCD99}     &  Staggered & $4.23(29)$      & $106(7)$    & $129(12)$    & $M_\rho$       \\
ALPHA-UKQCD 1999 \cite{MqALPHA99} & O(a) SW    &                 & $ 97(4)$    &              & $f_K$          \\
RIKEN-BNL 1999 \cite{MqBNL99}     & DWF        &                 & $ 95(26)$   &              & $f_\pi$        \\
QCDSF   1999 \cite{MqQCDSF99}     & O(a) SW    & $4.4(2)$        & $105(4)$    &              & $R_0$          \\
QCDSF   1999 \cite{MqQCDSF99}     & Wilson     & $3.8(6)$        & $87(15)$    &              & $R_0$          \\
\hline
\end{tabular}
\caption{Recent lattice estimates in MeV of $\bar m$ and $m_s$, 
in $\MSbar$ scheme at 2 GeV. SW stands for the Sheikhholeslami-Wohlert action.}
\label{t:mq}
\end{center}
\end{table}

First, the difference between $m_s(M_K)$ and $m_s(M_\phi)$, and the
variation with the observable used to fix $1/a$ are both symptoms of
the quenched approximation.  The data suggests that it is a $\sim
10\%$ effect, and at present constitutes the biggest
uncertainty. Second, the consistency of the results using different
actions (from Wilson to domain wall fermions), analyzed using the same
states and after an $a=0$ extrapolation to remove discretization
errors, shows that the lattice technology is robust and that we have
control over discretization errors.  Third, there has been much debate
over which renormalization constants (tadpole improved perturbative or
from the various non-perturbation methods) to use.  The data show that
after extrapolation to $a=0$, the difference is at most a few percent.
So the bottom line is that we now have many different methods and consistency
checks within the lattice approach for calculating light quark masses
and just need the computer power to shed the last approximation -- the
quenched approximation -- to get reliable estimates.

The only unquenched data (albeit for 2 degenerate dynamical flavors)
of the modern era (using improved action, nonperturbative
renormalization constants, and extrapolation to $a=0$) are the
preliminary results by the CPPACS collaboration. T. Kaneko at LATTICE 99 reported 
\begin{eqnarray}
(m_u + m_d)/2 &=&  3.3(4)  {\rm MeV} \nonumber \\
m_s(M_K)      &=&  84(7) {\rm MeV} \nonumber \\
m_s(M_\phi)   &=&  87(11) {\rm MeV} \nonumber \\
\end{eqnarray}
It is quite remarkable that $m_s(M_K)$ and $m_s(M_\phi)$ already show consistency. 
Also, the associated mass splittings in the baryon octet and decuplet are much 
improved. On the strength of these consistency checks I propose using 
\begin{equation}
m_s(\overline{MS}, \mu=2 GeV) = 85(10) {\rm MeV} \,.
\end{equation}

\section{$\Delta I = 3/2$ Electroweak Penguins: $B_7$ and $B_8$}

Current lattice calculations of the $\Delta I = 3/2$ amplitude rely on
CPT to relate $K \to \pi \pi$ to $K \to \pi$ with degenerate $K$ and
$\pi$.  Under these approximations, the calculation of $B_7$ and $B_8$
is equivalent to that of $B_K$ in complexity. Initially, there was a
problem of much larger 1-loop renormalization constants. This is now
under much better control through the development of better operators
and non-perturbative methods. A summary of results is given in
Table~\ref{t:B7B8}. All results using perturbative $Z$'s are
consistent, confirming that the calculation of the raw correlation
functions is under control. The APE calculation~\cite{BKB7B897APE}
using non-pertubative $Z$'s gives a value higher by $\sim
20\%$. However, since almost all the calculations have been done at
only one coupling, and anticipating that the extrapolation to the
$a=0$ limit will also be different for the two ways of estimating the
$Z$'s, it is too early to choose between the two values.  Calculations
at other values of the coupling are in progress and I anticipate we
will reduce the uncertainty to $< 10\%$ within the year.

The more important issue is whether tree-level CPT is sufficient to
relate $K \to \pi \pi$ to $K \to \pi$ with $M_K = M_\pi$. Since a
similar situation in $B_4$ suggests not~\cite{golterman}, Golterman
and Pallante are doing the needed 1-loop calculations. Thereafter, one
has to confront issues of removing the quenched approximation and
developing the technology for dealing with the physical case of $K \to
\pi \pi$ away from threshhold in case 1-loop CPT corrections are very
large. Sheer optimism propels me to believe that we will see progress
towards realistic estimates of these parameters in the next couple of years.

\begin{table}
\begin{center}
\medskip\noindent
\begin{tabular}{|l|c|c|c|c|}
\hline
    Fermion Action                  &  $Z$           &  $\beta$      &   $B^{3/2}_7$   & $B^{3/2}_8$ \\
\hline							              
Staggered \cite{B97stag}            & 1-loop TI      &  $6.0,\ 6.2^*$&   $0.62(3)(6)$  & $  0.77(4)(4)$ \\
Wilson    \cite{B97wilnLANL}        & 1-loop TI      &  $6.0$        &   $0.58(2)(7)$  & $  0.81(3)(3)$ \\
SW     \cite{BKB7B897APE}           & 1-loop BPT     &  $6.0$        &   $0.58(2)   $  & $  0.83(2)   $ \\
SW     \cite{BKB7B897APE}           & Non-pert.      &  $6.0$        &   $0.72(5)   $  & $  1.03(3)   $ . \\
TI SW  \cite{B98lellouch}           & 1-loop TI      &  $6.0, 6.2$   &   $0.72^{+5+2}_{-4-8} $  & $  0.80^{+8+1}_{-8-4}$ \\
\hline
\end{tabular}
\caption{Lattice estimates of $B_7$, and $B_8$ $ (NDR,\mu=2\GeV)$ for 
the amplitude $K \to \pi$. An asterisk implies extrapolation to $a=0$.}
\label{t:B7B8}
\smallskip
\end{center}
\end{table}

\section{Strong Penguin: $B_6$}

Lattice QCD does not yet have an estimate for $B_6$. In addition to
the issue of using CPT to relate $K \to \pi \pi$ to $K \to \pi$, there
is also the problem of mixing with lower dimension operators and large
renormalization factors. There are two calculations underway. One
using domain wall fermions as already discussed by Blum; and the
second by Kilcup and Pekurovsky using staggered
fermions~\cite{B6kilcup}. Kilcup \etal\ show that all the needed
correlation functions can be calculated with small statistical errors,
however, since the 1-loop $Z's$ for staggered fermions are very large
($\sim 100\%$) there are no reliable predictions. One definitely needs
non-perturbative calculation of these. It is too early to tell if
domain wall fermions will prove to be the method of choice.  In short,
$B_6$, which is crucial to understanding both the $\Delta I = 1/2$
rule and $\epsilon'$, is still an open problem in lattice QCD.

\vspace{.2in}
\noindent
\underline{Acknowledgements}

\vspace{.1in}
\noindent
It is a pleasure to thank Bruce and Jon and for organizing such a wonderful 
conference, and T. Bhattacharya, C. Bernard, and S. Sharpe for their 
comments.


\ifx\href\undefined\def\href#1#2{{#2}}\fi
\def\spireshome{http://www.slac.stanford.edu/cgi-bin/spiface/find/hep/www?FORMAT=WWW&}
\def\xxxhome{http://xxx.lanl.gov/abs/}
{\catcode`\%=12
\xdef\spiresjournal#1#2#3{\noexpand\protect\noexpand\href{\spireshome
                          rawcmd=find+journal+#1%2C+#2%2C+#3}}
\xdef\spireseprint#1#2{\noexpand\protect\noexpand\href{\spireshome rawcmd=find+eprint+#1%2F#2}}
\xdef\spiresreport#1{\noexpand\protect\noexpand\href{\spireshome rawcmd=find+rept+#1}}
\xdef\spireskey#1{\noexpand\protect\noexpand\href{\spireshome key=#1}}
\xdef\xxxeprint#1{\noexpand\protect\noexpand\href{\xxxhome #1}}
}
\def\eprint#1#2{\xxxeprint{#1/#2}{#1/#2}}
\def\report#1{\spiresreport{#1}{#1}}
\def\nohref{}

\makeatletter
\def\putpaper{\@ifnextchar [{\@putpaper}{\@putpaper[]}}
\def\@putpaper[#1]{\edef\refpage{\the\count0}%
              \def\nohref{}%
              {\def\ {+}\def\nohref##1{}\edef\temp{\ifx\relax#1\relax
               \noexpand\spiresjournal{\journalname}{\volume}{\refpage}%
               \else\noexpand\xxxeprint{#1}\fi}\expandafter}\temp
               {\sfcode`\.=1000{\journalname} \journalformat}\egroup}
\def\putpage{\@ifnextchar [{\@putpage}{\@putpage[]}}
\def\@putpage[#1]{\edef\refpage{\the\count0}%
              \def\nohref{}%
              {\def\ {+}\def\nohref##1{}\edef\temp{\ifx\relax#1\relax
	       \noexpand\spiresjournal{\journalname}{\volume}{\refpage}%
               \else\noexpand\xxxeprint{#1}\fi}\expandafter}\temp
              {\refpage}\egroup}
\def\dojournal#1#2 (#3){\def\journalname{#1}\def\volume{#2}%
                         \def\refyear{#3}\afterassignment\putpaper\bgroup
                         \count0=}
\def\morepage{\afterassignment\putpage\bgroup\count0=}
\def\supresslink{\def\spiresjournal##1##2##3{}}

\def\APNY#1{\dojournal{Ann.\ Phys.\ \nohref{(N.\ Y.)}}{#1}}
\def\CMP#1{\dojournal{Comm.\ Math.\ Phys.}{#1}}
\def\IJMPC#1{\dojournal{Int.\ J.\ Mod.\ Phys.}{C#1}}
\def\IJMPE#1{\dojournal{Int.\ J.\ Mod.\ Phys.}{E#1}}
\def\JAP#1{\dojournal{J.\ App.\ Phys.}{#1}}

\def\MPA#1{\dojournal{Mod.\ Phys.\ Lett.}{A#1}}
\def\MPLA#1{\dojournal{Mod.\ Phys.\ Lett.}{A#1}}
\def\NP#1{\dojournal{Nucl.\ Phys.}{B#1}}
\def\NPA#1{\dojournal{Nucl.\ Phys.}{A#1}}
\def\NPB#1{\dojournal{Nucl.\ Phys.}{B#1}}
\def\NPBPS#1{\dojournal{Nucl.\ Phys.\ \nohref(Proc.\ Suppl.\nohref)}{\nohref B#1}}
\def\NPAPS#1{\dojournal{Nucl.\ Phys.\ \nohref(Proc.\ Suppl.\nohref)}{\nohref A#1}}
\def\NC#1{\dojournal{Nuovo Cimento }{#1}}
\def\PRL#1{\dojournal{Phys.\ Rev.\ Lett.}{#1}}
\def\PR#1{\dojournal{Phys.\ Rev.}{#1}}
\def\PRep#1{\dojournal{Phys.\ Rep.}{#1}}
\def\PRB#1{\dojournal{Phys.\ Rev.}{B#1}}
\def\PRC#1{\dojournal{Phys.\ Rev.}{C#1}}
\def\PRD#1{\dojournal{Phys.\ Rev.}{D#1}}
\def\PRE#1{\dojournal{Phys.\ Rev.}{E#1}}
\def\PL#1{\dojournal{Phys.\ Lett.}{#1B}}
\def\PLA#1{\dojournal{Phys.\ Lett.}{#1A}}
\def\PLB#1{\dojournal{Phys.\ Lett.}{#1B}}
\def\RMP#1{\dojournal{Rev.\ Mod.\ Phys.}{#1}}
\def\PREP#1{\dojournal{Phys.\ Rep.}{#1}}
\def\ZEITC#1{\dojournal{Z.\ Phys.}{C#1}}
\def\ZPC#1{\dojournal{Z.\ Phys.}{C#1}}

\def\ie{{\sl i.e.}}
\def\etc{{\it etc.}}
\def\ibid{{\it ibid}}


\let\super=^
\catcode`\^=13
\def^{\ifmmode\super\else\initialsep\fi}

\def\journalformat{{\bf \volume}, \refpage\ (\refyear)}
\def\initialsep{}

{}

\end{document}